# Subwavelength Plasmonic Lattice Solitons in Arrays of Metallic Nanowires

F. Ye,[1] D. Mihalache,[2] B. Hu,[1,3] and N. C. Panoiu[4]

[1]*Department of Physics, Centre for Nonlinear Studies,
and The Beijing-Hong Kong-Singapore Joint Centre for Nonlinear and Complex Systems (Hong Kong),
Hong Kong Baptist University, Kowloon Tong, Hong Kong, China*
[2]*"Horia Hulubei" National Institute for Physics and Nuclear Engineering, Department of Theoretical Physics,
407 Atomistilor, Magurele-Bucharest, 077125, Romania*
[3]*Department of Physics, University of Houston, Houston, Texas 77204, USA*
[4]*Department of Electronic and Electrical Engineering, University College London,
Torrington Place, London WC1E 7JE, United Kingdom*

We predict theoretically that stable subwavelength plasmonic lattice solitons (PLSs) are formed in arrays of metallic nanowires embedded in a nonlinear medium. The tight confinement of the guiding modes of the metallic nanowires, combined with the strong nonlinearity induced by the enhanced field at the metal surface, provide the main physical mechanisms for balancing the wave diffraction and the formation of PLSs. As the conditions required for the formation of PLSs are satisfied in a variety of plasmonic systems, we expect these nonlinear modes to have important applications to subwavelength nanophotonics. In particular, we show that the subwavelength PLSs can be used to optically manipulate with nanometer accuracy the power flow in ultracompact photonic systems.

The downscaling of photonic devices for confining and manipulating optical energy at the nanoscale is one of the major challenges of nanophotonics [1]. When the size of conventional optical circuits is reduced to nanoscale, the spatial confinement of light is inherently limited by diffraction. One effective approach to overcome this limitation is to use surface plasmon polaritons (SPPs) [2,3]. In particular, by using SPP modes of metallic nanowires [4], chains of resonantly coupled metallic nanoparticles [5,6], tapered plasmonic waveguides [7], or cylindrical metallic gratings [8] one can spatially confine and guide optical energy over distances much smaller than the wavelength. These basic guiding nanostructures can be assembled in more complex plasmonic systems, such as Y splitters, Mach-Zehnder interferometers, and waveguide-ring resonators [9]. Despite these promising developments, there remains a basic challenge that one has yet to overcome in order to fully exploit the technological potential of plasmonic devices: they must provide the critical functionality of all-optic active control of light at nanoscale. Because of the strong enhancement of the field induced by the excitation of SPPs, and, consequently, the increased optical nonlinearity, SPPs are particularly suited for providing this functionality. While basic nonlinear optical processes have been demonstrated in a variety of plasmonic nanostructures, e.g., optical limiting and self-phase modulation in chains of structured nanoparticles [10] or second-harmonic generation in nanostructured metallic films [11,12], the physical constraints imposed by large in-plane extent of the optical field and out-of-plane operation of some of these devices preclude their integration in ultracompact plasmonic systems.

In this Letter, we present a very promising approach to achieve subwavelength confinement of the optical field guided by plasmonic nanostructures. In the proposed plasmonic nanostructure, which consists of an array of closely spaced parallel metallic nanowires embedded in a nonlinear optical medium (see Fig. 1), the optical nonlinearity induced by the field of the guiding modes of the nanowires compensates the discrete diffraction due to the optical coupling among the nanowires. As a result, nonlinear collective modes, which we call *plasmonic lattice solitons* (PLSs), are formed in the plasmonic array. Because the radius $a$ of the nanowires and their separation distance, $d$, are much smaller than the operating wavelength, $\lambda$, the spatial width of the PLSs can be significantly smaller than $\lambda$. Importantly, this remarkable property of PLSs cannot be achieved by using dielectric waveguide arrays, which also support discrete solitons [13–17], as the transverse size of such waveguides is comparable or larger than the wavelength.

Our analysis of the dynamics of PLSs is based on an extension to the nonlinear case of a coupled-mode theory, which captures the full vectorial character of the propagating modes of the metallic nanowires [18]. This fully vectorial description of the PLSs is essential for a rigorous analysis of their physical properties since the electric field of the modes of metallic nanowires has a large longitudinal component and therefore it cannot be described by a scalar function. Moreover, our analysis of the PLSs applies to the general case of a complex dielectric constant and as such it fully accounts for the losses in the nanowires. In particular, we use the Drude model for the dielectric constant of the metal, $\epsilon_m(\omega) = 1 - \frac{\omega_p^2}{\omega(\omega+i\nu)}$, and consider that the nano-wires are made of Ag ($\omega_p = 13.7 \times 10^{15}$ rad/s and $\nu = 2.7 \times 10^{13}$ rad/s [19]).

We start our analysis of the PLSs by expanding the total electric field $\mathbf{E}(\mathbf{r})$ and magnetic field $\mathbf{H}(\mathbf{r})$ in a superposition of the modes of a single nanowire, $\mathbf{E}_\perp(\mathbf{r}_\perp, z) = \sum_n \frac{a_n(z)}{\sqrt{P_n}} \mathbf{e}_\perp^{(n)}(\mathbf{r}_\perp)$, $E_z(\mathbf{r}_\perp, z) = \sum_n \frac{a_n(z)}{\sqrt{P_n}} \frac{\epsilon^{(n)}(\mathbf{r}_\perp)}{\epsilon(\mathbf{r}_\perp)} e_z^{(n)}(\mathbf{r}_\perp)$, and $\mathbf{H}(\mathbf{r}_\perp, z) = \sum_n \frac{a_n(z)}{\sqrt{P_n}} \mathbf{h}_\perp^{(n)}(\mathbf{r}_\perp)$, where $a_n$ is the mode amplitude in the $n$th nanowire, $\epsilon(\mathbf{r}_\perp)$ and $\epsilon^{(n)}(\mathbf{r}_\perp)$ are the dielectric constant of the plasmonic array and of an isolated nanowire, respectively, and $[\mathbf{e}^{(n)}(\mathbf{r}_\perp), \mathbf{h}^{(n)}(\mathbf{r}_\perp)]$ are the modal fields. These modes are normalized such that $P_n = \frac{1}{4}\int_S [\mathbf{e}^{(n)} \times \mathbf{h}^{(n)*} + \mathbf{e}^{(n)*} \times \mathbf{h}^{(n)}] \cdot \hat{\mathbf{z}} dS$ is the mode power. For simplicity, we assumed that the nanowires have only the fundamental TM mode ($h_z = 0$), whose nonvanishing field components, $e_r$, $e_z$, and $h_\phi$, depend only on the radial coordinate, $r_\perp$. These field components can be found analytically by solving the Maxwell equations, while the dispersion relation (the dependence of the complex propagation constant $\beta = \beta_r + i\beta_i$ on $\omega$) can be determined by imposing continuity conditions on the tangent fields at the metal-dielectric interface. As illustrated in Fig. 1(a), the large dielectric constant of metals, combined with the subwavelength transverse size of the nanowire, leads to a strong dependence of $\beta$ on the wavelength.

To find the mode amplitudes, we start from the unconjugated form of the Lorentz reciprocity theorem [20],

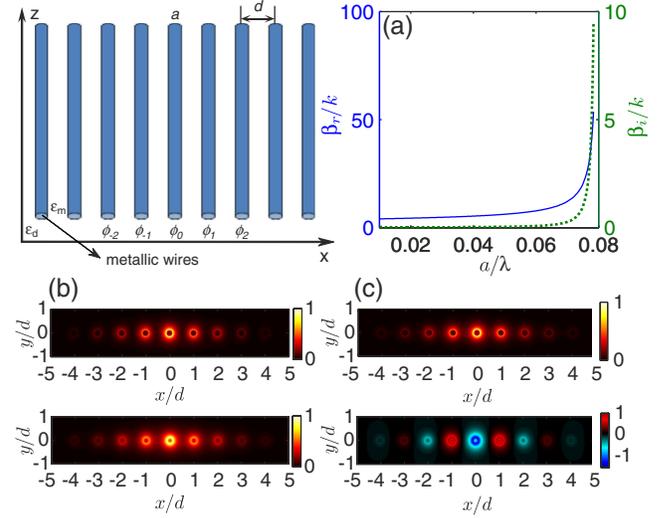

FIG. 1 (color online). Schematics of a 1D array of metallic nanowires with radius $a$ and separation distance $d$ (top left). (a) Real and imaginary parts of the propagation constant of the fundamental TM mode, $\beta_r$ and $\beta_i$, respectively. Panels (b) and (c) show the transverse profile of the amplitude (top) and longitudinal component (bottom) of the electric field of unstaggered and staggered PLSs, respectively, for $\lambda = 1550$ nm and $n_b = 3.5$. In (b) and (c), $\delta n_{nl} = -0.05$ ($n_2 = -4 \times 10^{-18}$ m$^2$/W) and $\delta n_{nl} = 0.05$ ($n_2 = 4 \times 10^{-18}$ m$^2$/W), respectively. The nanowires have $a = 40$ nm and $d = 8a$.

$$\frac{\partial}{\partial z}\int_S [\mathbf{E}_1(\mathbf{r}, \omega) \times \mathbf{H}_2(\mathbf{r}, \omega) - \mathbf{E}_2(\mathbf{r}, \omega) \times \mathbf{H}_1(\mathbf{r}, \omega)] \cdot \hat{\mathbf{z}} dS = i\omega \int_S [\epsilon_2(\mathbf{r}) - \epsilon_1(\mathbf{r})]\mathbf{E}_1(\mathbf{r}, \omega) \cdot \mathbf{E}_2(\mathbf{r}, \omega) dS, \quad (1)$$

where $(\mathbf{E}_1, \mathbf{H}_1)$ and $(\mathbf{E}_2, \mathbf{H}_2)$ are solutions of the Maxwell equations corresponding to the dielectric constants $\epsilon_1(\mathbf{r})$ and $\epsilon_2(\mathbf{r})$, respectively. If we choose $(\mathbf{E}_1, \mathbf{H}_1)$ and $(\mathbf{E}_2, \mathbf{H}_2)$ to be the fields in the plasmonic array and the fields of a backward ($-z$) propagating mode in the $n$th nanowire, respectively, and $\epsilon_1(\mathbf{r}) = \epsilon(\mathbf{r}_\perp) + \delta\epsilon_{nl}(\mathbf{r})$ and $\epsilon_2(\mathbf{r}) = \epsilon^{(n)}(\mathbf{r}_\perp)$ the corresponding dielectric constants (here $\delta\epsilon_{nl}$ is the nonlinear change in the dielectric constant), the Eq. (1) leads to the following system of coupled equations describing the mode amplitudes $a_n(z)$:

$$\left(i\frac{d}{dz} + \beta\right)\left(a_n + \frac{c_{n,n-1} + c_{n-1,n}}{2c_{nn}}a_{n-1} + \frac{c_{n,n+1} + c_{n+1,n}}{2c_{nn}}a_{n+1}\right) = \frac{1}{c_{nn}}(K_{nn}a_n + K_{n,n-1}a_{n-1} + K_{n,n+1}a_{n+1}) + \frac{\Gamma_{nn}}{c_{nn}}|a_n|^2 a_n. \quad (2)$$

In these equations $c_{nm} = \frac{1}{2\sqrt{P_n P_m}}\int_S [\mathbf{e}^{(n)} \times \mathbf{h}^{(m)}] \cdot \hat{\mathbf{z}} dS$, $K_{nm} = \frac{\omega}{4\sqrt{P_n P_m}}\int_S \Delta\epsilon^{(n)}(\mathbf{r}_\perp)F_{nm}(\mathbf{r}_\perp)dS$, where $\Delta\epsilon^{(n)}(\mathbf{r}_\perp) = \epsilon^{(n)}(\mathbf{r}_\perp) - \epsilon(\mathbf{r}_\perp)$ and $F_{nm}(\mathbf{r}_\perp) = \mathbf{e}_\perp^{(n)} \cdot \mathbf{e}_\perp^{(m)} - \frac{\epsilon^{(m)}}{\epsilon}e_z^{(n)}e_z^{(m)}$, and the nonlinear coefficient $\Gamma_{nn} = -\frac{\epsilon_0 n_b \omega n_2}{2P_n^2} \times \int_S \alpha_n(\mathbf{r}_\perp)F_{nn}(\mathbf{r}_\perp)dS$, where $n_b$ is the refractive index of the background, $n_2$ is the Kerr coefficient, and $\alpha_n(\mathbf{r}_\perp) = |\mathbf{e}_\perp^{(n)}|^2 + |\frac{\epsilon^{(n)}}{\epsilon}e_z^{(n)}|^2$. Note that in deriving Eq. (2) we have neglected the nonlinear interaction among the nanowires, i.e., $\delta\epsilon_{nl}(\mathbf{r}) \approx 2\epsilon_0 n_b n_2 \sum_n \frac{\alpha_n(\mathbf{r}_\perp)}{P_n}|a_n(z)|^2$.

Our calculations show that if $d$ is of the order of a few hundred nanometers $(c_{n,n\pm 1} + c_{n\pm 1,n})/c_{nn} < 1\%$ and thus the corresponding terms in Eq. (2) can be neglected. Moreover, by rescaling the mode amplitudes, $a_n(z) = \sqrt{P_0}\phi_n(z)\exp[i(\beta - K_{nn}/c_{nn})z]$, with $P_0$ the power in the zeroth nanowire, the system (2) can be simplified as

$$i\frac{d\phi_n}{dz} + \kappa(\phi_{n-1} + \phi_{n+1}) + \gamma|\phi_n|^2\phi_n = 0, \quad (3)$$

where $\kappa = -K_{n,n\pm 1}/c_{nn}$ and $\gamma = -P_0\Gamma_{nn}/c_{nn}$. This is the discrete nonlinear Schrödinger equation, which is known to have soliton solutions [13]. We emphasize that for our plasmonic array $\kappa < 0$, so that the linear dispersion relation, $k_z = 2\kappa\cos(k_x d)$, implies that anomalous (normal) diffraction occurs at $k_x = 0$ ($k_x = \pi/d$), which is opposite to the case of dielectric waveguide arrays.

The soliton solutions of Eq. (3) are sought in the form $\phi_n(z) = u_n \exp(i\rho z)$, where the amplitudes $u_n$ are independent of $z$ and $\rho$ is the soliton wave number. We found that our plasmonic array supports two types of PLSs, unstaggered and staggered solitons. In the case of unstaggered (staggered) PLSs the phase difference of the mode

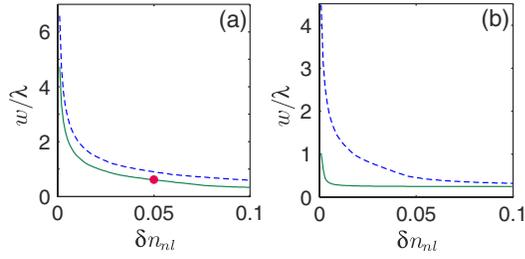

FIG. 2 (color online). Soliton width $w$ vs the nonlinear index change $\delta n_{nl}$, calculated for $a = 40$ nm and (a) $\lambda = 1550$ nm, $d = 8a$, and (b) $\lambda = 632$ nm, $d = 4a$. The solid and dashed lines correspond to $n_b = 3.5$, $n_2 = 4 \times 10^{-18}$ m$^2$/W (Si) and $n_b = 2.8$, $n_2 = 1.4 \times 10^{-18}$ m$^2$/W (As$_2$Se$_3$), respectively. The dot in (a) corresponds to the soliton in Fig. 1(c).

amplitudes in adjacent nanowires is equal to zero ($\pi$). The spatial profile of the amplitude and longitudinal field of unstaggered (staggered) PLSs corresponding to a nonlinear index change of $\delta n_{nl} = -0.05$ ($\delta n_{nl} = 0.05$), with $\delta n_{nl} = 2\delta \epsilon_{nl}/(\epsilon_0 n_b)$, are shown in Fig. 1(b) [Fig. 1(c)]. Note that due to the inverted linear dispersion relation, staggered (unstaggered) solitons are formed in self-focusing (self-defocusing) media, which is opposite to the case of dielectric waveguide arrays [17]. Importantly, the soliton full width at half maximum is $w \approx 0.6\lambda$, i.e., the soliton has subwavelength extent. The dependence of the soliton width on $\delta n_{nl}$ is presented in Fig. 2. As expected, this figure shows that the soliton width decreases with the strength of the induced nonlinearity and increases with the wavelength.

Figure 3 shows the propagation of a staggered PLS in the plasmonic array. It can be seen that in the lossless ($\nu = 0$) linear propagation regime ($n_2 = 0$), the plasmon field experiences significant discrete diffraction [see Fig. 3(a)]. However, when the optical nonlinearity is taken into account, the plasmon field maintains its shape during propagation, which means that a PLS is formed. When the optical losses are included, the absorption coefficient is $2\Im(\rho) = 910$ cm$^{-1}$, which corresponds to a decay length of 11 $\mu$m. On the other hand, Fig. 3(c) illustrates that when both optical losses and the nonlinearity are included, the plasmon field of the PLS retains its initial width over a propagation distance of $\sim 20$ $\mu$m. Thus, an experimental observation of subwavelength PLSs can be realized even without a gain medium. A solitonlike propagation requires a gain of $g = 910$ cm$^{-1}$, which can be easily achieved in a practical experimental setting [21].

In a common experimental setting, solitons are excited from input Gaussian optical beams, which do not have the exact profile of the actual soliton. We therefore investigate the excitation of PLSs from Gaussian beams whose width and amplitude are optimized so as to lead to the shortest *soliton formation length*, which is the distance required for a beam to reshape itself into a soliton. The generic scenario of soliton formation is illustrated in Fig. 4. It shows that during a propagation distance of just a few tens of micrometers the input beam sheds off part of its energy as radiative waves, the remaining plasmon field evolving into the PLS. During this latter transient stage the width

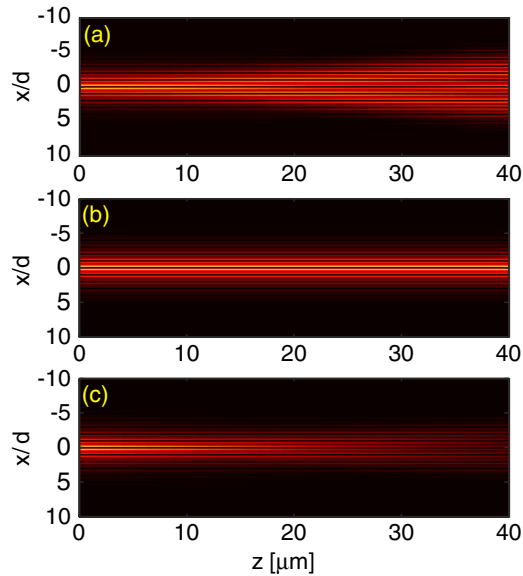

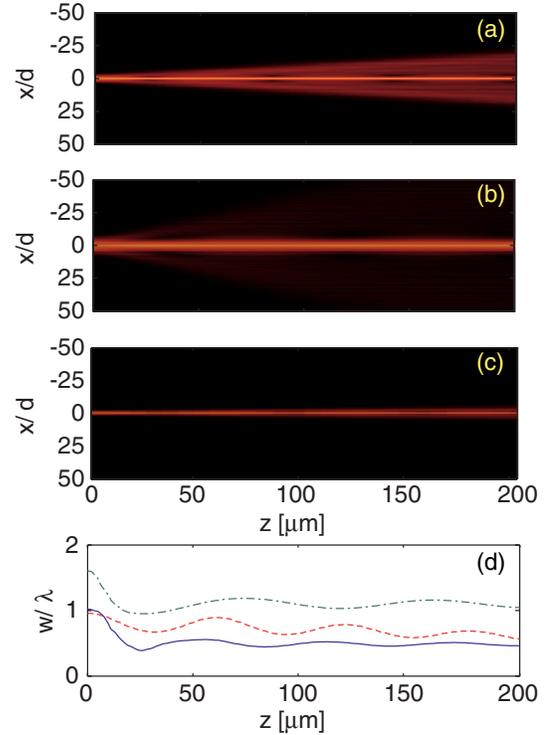

FIG. 3 (color online). Linear (a) and nonlinear (b) propagation of the PLS in Fig. 1(c) in the lossless plasmonic array with $n_b = 3.5$, $n_2 = 4 \times 10^{-18}$ m$^2$/W, $\lambda = 1550$ nm, $a = 40$ nm, and $d = 8a$. (c) propagation of the same PLS in the lossy plasmonic array.

FIG. 4 (color online). Propagation of a Gaussian beam with $\lambda = 1550$ nm in a plasmonic array with (a) $d = 6a$, (b) $d = 4a$, and (c) $d = 8a$. (d) The beam width vs $z$: the solid, dash-dotted, and dashed curves correspond to panels (a),(b), and (c), respectively.

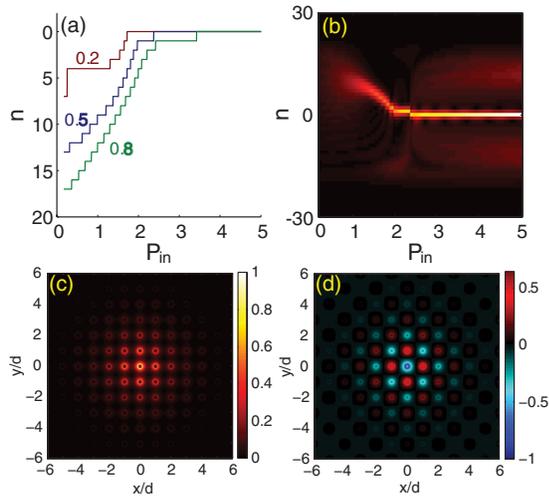

FIG. 5 (color online). (a) The location of the peak amplitude at the output facet of a plasmonic array with length of 150 $\mu$m, $a = 0.03\lambda$, and $d = 8a$ vs $P_{in}$, determined for different phase tilt $k_0$ ($\lambda = 1550$ nm). (b) The amplitude distribution of the output plasmon field vs $P_{in}$, calculated for $k_0 = 0.5$. In (c) and (d), the normalized transverse and longitudinal electric field of a staggered 2D PLS, respectively, calculated for $a = 40$ nm, $d = 8a$, and $\lambda = 1550$ nm. The background material is Si.

of the beam presents a damped oscillatory evolution [see Fig. 4(d)]. Importantly, Fig. 4(d) also shows that the soliton formation length is strongly dependent on the geometry of the plasmonic array. For example, for $a = 40$ nm and $\lambda = 1550$ nm, the shortest soliton formation length, of about 50 $\mu$m, is achieved for $d = 6a$.

In order to illustrate the technological potential of PLSs, e.g., to subwavelength chip-level active nanodevices, we show that the dynamics of PLSs in the plasmonic array can be easily controlled *via* optical means. To this end, we launched into the plasmonic array a staggered PLS with an initial phase tilt, $\phi_n(0) = u_n \exp(ik_0 x)$, with $k_0$ being the transverse wave number. Figure 5 presents the power dependence of the dynamics of the PLS in the plasmonic array. Thus, at low input power, $P_{in}$, the PLS moves across the array, the transverse shift increasing with $k_0$. However, as $P_{in}$ increases, the transverse shift of the plasmon field decreases, and, finally, for $P_{in}$ exceeding a certain threshold value, the PLS is trapped at its initial location. Although a similar soliton dynamics have been observed in dielectric waveguide arrays [17], PLSs provide the critical functionality of all-optical control *with subwavelength precision* of the spatial confinement of the optical field (note that $d < \lambda/4$ for the plasmonic array in Fig. 5).

It should be noted that unlike the plasmon solitons in layered metallo-dielectric structures [22,23], the PLSs discussed here can readily be extended to 2D geometries, in which case new types of nonlinear plasmonic modes, such as discrete plasmonic vortex solitons, should exist. As shown in Figs. 5(c) and 5(d), PLSs can form in 2D plasmonic arrays, their size being subwavelength in this case, too ($w = 0.72\lambda$). Furthermore, the existence of subwavelength PLSs is not limited to arrays of metallic nanowires. They can be excited in any system of coupled plasmonic waveguides, as long as the transverse dimension of the waveguides is much smaller than the wavelength. Thus, one can consider arrays made of coupled wedge waveguides [9] or coupled chains of interacting metallic nanoparticles [10]. Moreover, the size of the plasmonic array, and implicitly the size of the PLSs, can be significantly reduced if one uses deeply scaled down nanostructures, such as metallic carbon nanotubes [24,25].

In summary, we have demonstrated theoretically that subwavelength PLSs are formed in 1D arrays of metallic nanowires embedded into a host dielectric medium with Kerr nonlinearity. The excitation of PLSs from Gaussian beams has also been investigated and their potential use to all-optical nanodevices has been discussed. We expect these results to enable exciting new developments in nanophotonics and subwavelength optics.

The work of N. C. P. has been supported by the EPSRC, Grant No. EP/G030502/1.